\documentclass[pdflatex,sn-mathphys-num]{sn-jnl}

\usepackage{graphicx}%
\usepackage{multirow}%
\usepackage{amsmath,amssymb,amsfonts}%
\usepackage{amsthm}%
\usepackage{mathrsfs}%
\usepackage[title]{appendix}%
\usepackage{xcolor}%
\usepackage{textcomp}%
\usepackage{manyfoot}%
\usepackage{booktabs}%
\usepackage{algorithm}%
\usepackage{algorithmicx}%
\usepackage{algpseudocode}%
\usepackage{listings}%

\theoremstyle{thmstyleone}%

\theoremstyle{thmstyletwo}%

\theoremstyle{thmstylethree}%

\begin{document}

\title[Conformal Risk Certificates for Portfolio Optimization]{Conformal Risk Certificates for Portfolio Optimization Under Uncertainty: A Gated Reinforcement-Learning Approach}

\author[1]{\fnm{Mohammad} \sur{Zarei}}\email{mohammad.zarei14@sharif.edu}

\author[2]{\fnm{Amirhossein} \sur{MohammadpourAzari}}\email{a.mohammadpour78@gmail.com}

\author*[1]{\fnm{Moslem} \sur{Habibi}}\email{mhabibi@sharif.edu}

\affil[1]{\orgdiv{Industrial Engineering Department}, \orgname{Sharif University of Technology}, \state{Tehran}, \country{Iran}}

\affil[2]{\orgdiv{Computer Engineering Department}, \orgname{Sharif University of Technology}, \state{Tehran}, \country{Iran}}

\abstract{Reinforcement-learning portfolio policies can optimize allocations from prices, covariances, and prior positions, yet they lack a calibrated, replayable mechanism for certifying that a proposed allocation respects a bounded risk envelope before execution. Regulators increasingly expect automated portfolio systems to provide exactly this kind of pre-trade audit trail, interpretable risk limit, and human-oversight mechanism.
Gated Conformal Semantic Risk Control addresses this gap by separating return-seeking allocation from text-conditioned downside-risk verification. An instruction-tuned language model maps timestamped disclosures into schema-validated semantic risk states over macro, liquidity, earnings, and sector channels. These states induce correlated ellipsoidal factor-shock uncertainty sets whose conformal radii are calibrated by rolling finite-sample prediction over localized semantic nonconformity buckets. An asymmetric second-order cone gate then checks each raw PPO proposal against a certified worst-case semantic loss budget. The gate stays inactive for compliant proposals and intervenes only when that budget is exceeded. In walk-forward out-of-sample tests on nine liquid U.S. Select Sector SPDR ETFs, Raw PPO earns a 12.7\% annual return while incurring a 4.05\% semantic violation rate. Gated-CSRC removes all observed out-of-sample semantic limit breaches while retaining an 11.7\% annual return. This 100-basis-point auditable safety premium is smaller than the yield loss imposed by unconditional hard projection, and exact McNemar testing confirms the compliance gain at $p < 10^{-15}$. Loughran--McDonald lexicon controls and time-shuffled text controls fail to reproduce this certificate.}


\maketitle

\section{Introduction}\label{sec:introduction}

Quantitative portfolio theory reads the market through prices, returns, and estimated moments, although the events that move portfolios often reach investors first as language. A regulatory inquiry, an earnings warning, a funding-market strain, or a geopolitical escalation enters the public record as text before it appears in a realized covariance matrix $\widehat{\Sigma}_t$ or a rolling volatility estimate. A policy that waits for prices to confirm the event commits capital against a risk geometry that the documents may already contradict. This delay does not reflect a weaker estimator alone. It follows from the order in which information arrives. Text can describe stress before price histories absorb it, which creates a pre-trade information deficit precisely when backward-looking risk estimates are least reliable.

The automation of portfolio allocation in capital markets has advanced faster than the control procedures used to audit it. Institutional regulators increasingly expect automated investment systems to provide pre-trade risk limits, replayable decision records, and human-interpretable oversight. The EU Artificial Intelligence Act requires human oversight measures for high-risk AI systems, including mechanisms that allow assigned natural persons to interpret outputs, monitor operation, and intervene when the system does not behave as intended \cite{euai2024act}. In portfolio management, SEC Rule 18f-4 places derivatives use within a formal risk-management program that includes risk identification, risk assessment, stress testing, board-facing oversight, and a VaR-based limit for leverage risk \cite{sec2020rule18f4}. These sources do not prescribe the method developed here, nor do they make a portfolio algorithm legally compliant by design. They clarify the direction of institutional demand. Automated allocation systems need separable risk control before execution, not only explanations after execution.

That deficit limits both classical portfolio control and modern reinforcement learning. Mean--variance selection formalized the risk--return trade-off, and Expected Shortfall shifted attention toward the tail losses that variance can smooth away \cite{markowitz1952portfolio,rockafellar2000optimization}. Direct and deep reinforcement learning extended this line by training policies to allocate capital through market interaction, transaction costs, and portfolio feedback rather than through a separate return forecast \cite{moody2001learning,jiang2017deep,zhang2022cost}. These systems remain disciplined because their objectives are numerical and testable. They also remain exposed because the state is usually price-bound. LLM-based trading systems move in the opposite direction. They ingest filings, news, and analyst language, yet often convert that language into asset rankings, directional calls, or scalar sentiment scores \cite{kirtac2024sentiment,choi2026text,kim2026sector}. The error is not computational. It is epistemological. A fluent model is not a calibrated model, and softmax confidence or rationale fluency does not provide empirical coverage over future factor shocks.

Gated Conformal Semantic Risk Control responds to this problem by separating proposal from execution. The reinforcement-learning actor proposes a raw target allocation $\tilde{w}_t$, but it receives no authority to execute unless the allocation satisfies a time-aligned semantic certificate. The language model does not forecast returns, rank assets, or select trades. It fills a schema-bound risk state extracted from documents available at decision time. That state defines a certified set of adverse factor shocks, and the executed allocation must remain feasible under that set. The architecture is therefore a neuro-symbolic pre-trade risk governor for reinforcement-learning portfolio allocation, with the learned policy treated as a proposal engine and the certificate treated as the executable control object.

The control object turns qualitative evidence into portfolio geometry. Time-aligned documents $D_t$ are mapped into a structured semantic state $z_t$, whose fields encode macroeconomic distress, liquidity strain, earnings pressure, sector exposure, and event severity. These fields determine a shock center $m_t$ and a positive semidefinite shape matrix $\Sigma_t$, while a rolling conformal procedure chooses a radius $\rho_t$ from past nonconformity scores within comparable semantic histories \cite{vovk2005algorithmic}. The resulting certificate
\[
C_t=(m_t,\Sigma_t,\rho_t,l_t)
\]
defines the adverse factor-shock set against which a proposed allocation is checked. The model proposes semantic direction and shape. Realized historical errors determine the radius of trust. A certificate with this form can be stored, replayed, and tested without appealing to the hidden activations of the language model.

Once the certificate exists, execution reduces to a gated feasibility decision. The raw policy output $\tilde{w}_t$ first faces a closed-form worst-case semantic loss evaluation over the ellipsoidal set induced by $C_t$, so the gate avoids nested adversarial search. If the allocation stays within the semantic loss budget $l_t$, it executes unchanged. If it breaches the budget, the system solves a second-order cone projection that finds the nearest feasible portfolio under semantic loss, leverage, turnover, volatility, sector, cash, and transaction-cost constraints. The intervention is asymmetric. Calm proposals are not penalized by needless projection, while certificate-breach days receive an explicit correction before capital is deployed \cite{wang2022risk,park2024twin}.

The paper makes three contributions. First, it formalizes semantic risk certificates that convert timestamped financial prose into conformally calibrated factor-shock uncertainty sets, so language enters portfolio control as a coverage-targeted risk object rather than as a forecast or sentiment score. Second, it develops a gated distribution-aware projection that separates policy proposal from execution and imposes pre-trade constraints only when the raw action breaches the current certificate. Third, it defines an auditable provenance protocol that links each executed allocation to the source documents, extracted semantic state, conformal bucket, certificate radius, raw action, projected action, intervention flag, and constraint residuals. This record makes the unit of trust a replayable decision trace rather than an explanation generated after the trade \cite{gonzalez2024portfolio,jadhav2025large}.

The empirical study runs on nine U.S. Select Sector SPDR ETFs over 2015--2023, with out-of-sample folds that include the 2020 crash. Raw PPO delivers strong financial performance but breaches the time-aligned certificate on $4.05\%$ of out-of-sample days on average across the annual test folds. Gated-CSRC reduces that rate to $0.00\%$, with an exact McNemar test giving $p<10^{-15}$, while Jobson--Korkie and Newey--West tests detect no statistically significant gap in Sharpe ratio or mean return. The annualized return falls from $12.7\%$ to $11.7\%$. The reported result is a measured safety trade-off, not unconstrained alpha dominance.

\section{Literature review}
\label{sec:literature_review}

Sequential portfolio choice, ambiguity-aware optimization, and financial language processing developed along separate lines. Dynamic policies learn from price histories, natural language models extract information from disclosures and news, and worst-case optimization translates uncertain inputs into feasible portfolio constraints. The unresolved interface is geometric. Existing systems rarely allow timestamped prose to alter the action set available to a continuous portfolio policy before execution. A policy may observe prices, a sentiment model may score documents, and an optimizer may guard against numerical uncertainty, yet the decision at time $t$ still lacks a calibrated map from textual evidence to portfolio feasibility. Gated-CSRC treats language as a source of auditable risk geometry rather than as a direct trading signal.

\subsection{Sequential portfolio decision-making and downside-risk control}

Feasibility first entered portfolio theory through static risk-return selection. \citet{markowitz1952portfolio} formalized diversification by separating expected return from variance, while Expected Shortfall shifted attention toward tail losses that variance can conceal \cite{rockafellar2000optimization}. Worst-case portfolio optimization extended this logic by replacing a single estimated distribution with ambiguity sets over returns, means, covariances, or distributional neighborhoods \cite{BenTalNemirovski1998,GoldfarbIyengar2003,BertsimasSim2004,DelageYe2010}. These methods give portfolio choice an explicit feasible region under uncertain inputs. Their informational boundary is equally clear. An ellipsoid built from past moments can bound numerical estimation error, whereas it cannot parse whether a central-bank statement, liquidity warning, or sanctions headline changes the conditional shape of sector losses before those losses enter realized covariance.

That static boundary motivated sequential trading systems that learn directly from market interaction. Recurrent reinforcement learning optimized financial performance functions under transaction costs and recurrent state dependence \cite{moody2001learning}. Deep portfolio policies later learned continuous allocations from price tensors, including architectures with portfolio-vector memory and online stochastic batches \cite{jiang2017deep}. Graph, convolutional, transformer, and permutation-invariant designs improved cross-asset representation, rebalancing behavior, and adaptation in larger universes \cite{SoleymaniPaquet2021,Marzban2023,ren2025time}. These systems show that neural policies can extract temporal and cross-sectional structure that static allocation cannot. They also inherit a closed state description. Market crashes appear in the training record as price paths, drawdowns, and volatilities, while the textual catalysts that preceded those paths often remain outside the policy input.

Risk-sensitive portfolio learning tried to repair that exposure from inside the training objective. Expected maximum drawdown, Calmar-ratio objectives, CVaR-aware policies, cost-sensitive rewards, uncertainty-aware agents, and constrained rebalancing systems penalize adverse trajectories or alter action selection under stressed states \cite{AlmahdiYang2017,AlmahdiYang2019,wang2022risk,zhang2022cost,Cui2023}. General constrained reinforcement learning gives a related formal precedent by specifying both rewards and constraints rather than forcing all behavior into a scalar reward \cite{Achiam2017}. Training-time risk control and execution-time feasibility remain different mechanisms. A penalty can change the average policy learned across years of data, but it need not reject a specific unsafe allocation on a particular trading day. Gated-CSRC therefore places a convex gate after the learned policy, so that $\widehat{w}_t$ becomes executable only after satisfying semantic loss, leverage, turnover, volatility, sector, and cash constraints.

\subsection{Financial NLP and the epistemic boundary of LLM trading agents}

The policy gate needs language because financial text carries information that price histories often absorb with delay. \citet{LoughranMcDonald2011} showed that general sentiment dictionaries misread financial documents, since domain-specific words can reverse ordinary affective meaning. Their dictionary remains a transparent baseline for disclosure tone. Its compression is severe. A document becomes a count vector or a scalar negativity score, which can register broad adverse tone but cannot express how funding stress in one sector may interact with leverage in another. Bag-of-words evidence lacks the relational memory needed to describe cross-sectional contagion or conditional factor-shock dispersion.

Contextual models expanded that evidence space. Media tone predicts market movements in earlier financial text studies, and domain-adapted transformers later improved sentiment extraction from financial documents \cite{Tetlock2007,Araci2019}. Subsequent work used financial language models, news embeddings, abnormal-return labels, and sector-conditioned reasoning to forecast assets or rank portfolios \cite{FinGPT2023,kirtac2024sentiment,choi2026text,kim2026sector}. This pathway treats language as semantic alpha prediction. It is useful for measuring whether text contains tradable information, but it also transfers an old allocation problem into a higher-dimensional representation. A noisy return forecast remains noisy when language produces it rather than prices. If an uncalibrated sentiment score or LLM ranking enters the optimizer as expected return, the portfolio can overfit semantic variance with the same fragility that affects numerical alpha forecasts.

The risk grows when language models receive direct trading authority. LLM-assisted margin systems and LLM-refined reinforcement learners use reasoning models to revise allocations, simulate trader judgment, or select market exposures \cite{gu2024adaptive,wang2025gems}. The relevant limitation is calibration. Neural confidence need not match empirical correctness, and modern networks can be poorly calibrated even when predictive accuracy appears high \cite{Guo2017}. A fluent macroeconomic rationale therefore does not provide a frequentist coverage statement for the losses that may follow a regime break. The same window that makes automated intervention attractive also weakens the reliability of softmax confidence and narrative plausibility.

Auditability imposes a sharper requirement than explanation. Post-hoc tools such as local surrogate explanations and Shapley attributions can identify influential features after a model has acted \cite{Ribeiro2016,lundberg2017unified}. Interpretable-model arguments also warn that explanation layers can mask the opacity of the underlying decision rule \cite{Rudin2019}. Portfolio execution needs a different object. A compliance officer must be able to reconstruct which document arrived before the trade, which semantic state was extracted, which uncertainty radius was calibrated, which constraints were active, and why the final allocation satisfied them. An attention map after a drawdown does not prove that the allocation respected a semantic loss envelope at time $t$.

\subsection{Distributional ambiguity control and conformal calibration}

The required audit object has a natural geometric form. Independent box constraints treat each factor shock as separable, although sectors often fail in correlated clusters during funding stress, policy shocks, or liquidity withdrawals. Ellipsoidal ambiguity sets can encode this joint structure through a shape matrix such as $\Sigma_t^{\mathrm{sem}}$, producing second-order cone constraints that bind cross-sectional stress rather than only marginal exposures. Classical ambiguity-set methods already show how moment uncertainty and covariance information can shape portfolio feasibility \cite{GoldfarbIyengar2003,DelageYe2010}. Gated-CSRC changes the source of the shape. Instead of estimating every direction from historical returns alone, it lets timestamped semantic risk states modify the cross-factor geometry that the proposed portfolio must survive.

A semantic shape matrix still requires scale. Conformal prediction supplies that empirical scale without assuming a parametric return distribution. The conformal framework constructs prediction or uncertainty sets from nonconformity scores and gives finite-sample coverage under exchangeability \cite{vovk2005algorithmic,ShaferVovk2008,AngelopoulosBates2021}. Financial time series strain global exchangeability because crises, policy regimes, and liquidity states change the distribution of future shocks. Gated-CSRC therefore calibrates scores within rolling semantic buckets rather than pooling all market histories into a single residual sample. This does not claim exact stationarity. It uses time-aligned semantic states to seek local comparability among calibration residuals, so that the radius $\rho_t$ reflects the current textual risk condition rather than an average of incompatible regimes.

\subsection{Regulatory oversight and auditability in AI-driven capital markets}

The same geometric object also has an institutional interpretation. The EU Artificial Intelligence Act requires human oversight for high-risk AI systems, and SEC Rule 18f-4 places derivatives use within a formal risk-management program for registered funds\cite{euai2024act,sec2020rule18f4}. These requirements make the distinction between post-hoc explanation and pre-trade certification economically meaningful: existing explanation tools can identify influential variables after a decision, but they cannot demonstrate that an executed allocation satisfied a semantic loss constraint before capital was deployed. The certificate formalized in Section 3 treats this institutional requirement as a boundary condition on the optimization problem, not as the paper's primary contribution.

\subsection{Synthesis and positioning of Gated-CSRC}

The literature leaves four correct ideas in tension. Tail loss must supplement variance. Continuous reinforcement learning can allocate dynamically in high-dimensional markets. Financial language contains information that prices may absorb with delay. Conformal and ambiguity-set methods can turn empirical uncertainty into feasible-set restrictions. Problems arise when these ideas collapse into one autonomous trading agent. Price-only RL omits semantic catalysts, semantic alpha models pass uncalibrated forecasts to optimizers, agentic LLMs confuse reasoning fluency with coverage, and classical ambiguity sets often remain tied to historical numerical moments. The missing contribution is an interface that assigns each component a limited authority.

Gated-CSRC implements that division of computational labor. The reinforcement learning policy proposes an alpha-seeking portfolio $\widehat{w}_t$ from market states. The open-source language model does not trade, rank assets, or forecast returns. It produces a schema-valid semantic risk state from timestamped text, leaving a replayable provenance record. Rolling conformal calibration converts realized factor-shock errors into an empirical radius. The convex gate then projects the proposal into a feasible portfolio $w_t$ under semantic loss, leverage, turnover, volatility, sector, and cash restrictions. Language influences execution only through the feasible action set, and every intervention can be traced from source document to calibrated constraint.

\section{Methodology}
\label{sec:methodology}

\begin{figure}[h]
    \centering
    \includegraphics[width=1.0\textwidth]{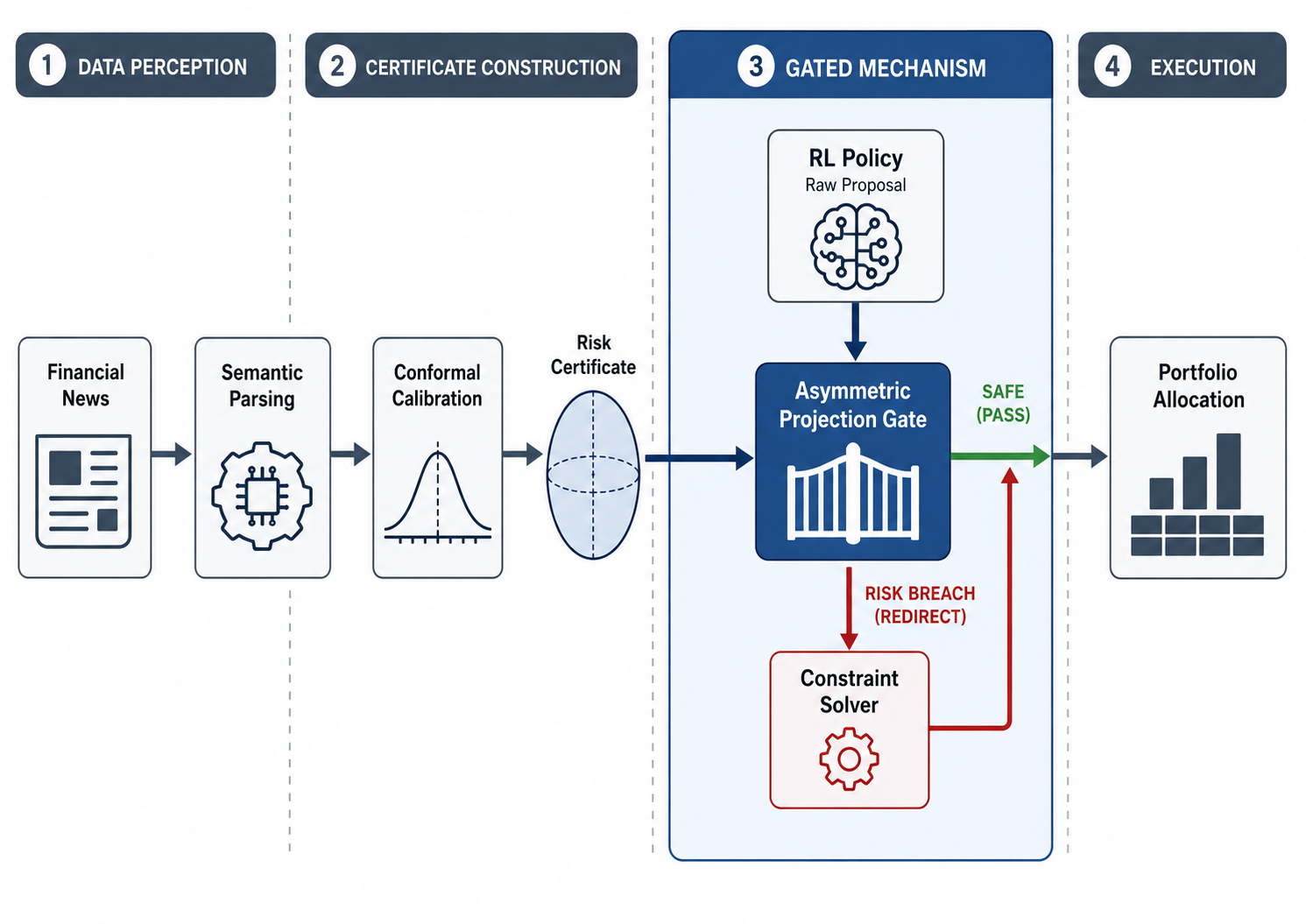}
    \caption{Gated-CSRC pre-trade execution framework. Timestamped market data and financial text are converted into semantic risk certificates, which define calibrated uncertainty sets and a semantic loss budget. The PPO actor proposes portfolio weights, while the gated projection layer executes the proposal only when the pre-trade certificate is satisfied.}
    \label{fig_gated_csrc_framework}
\end{figure}

Gated Conformal Semantic Risk Control separates portfolio proposal from portfolio execution. Price histories and timestamped documents first define the observable market state. A schema-bound language model output then becomes a semantic risk certificate over risky sector returns. The reinforcement-learning actor proposes a raw allocation, while a gated conic layer executes the proposal only after a pre-trade loss check. Figure~\ref{fig_gated_csrc_framework} displays the four-stage pipeline from data perception to certified execution.

\subsection{Portfolio setting}
\label{subsec_portfolio_setting}

The investment universe contains $N$ risky Select Sector SPDR ETFs indexed by $\mathcal{A}=\{1,\ldots,N\}$ and one cash instrument indexed by zero. At trading date $t$, the portfolio is $w_t=(w_{0,t},w_{1,t},\ldots,w_{N,t})^\top\in\mathbb{R}^{N+1}$, where $w_{0,t}$ denotes cash. Feasible allocations lie on the long-only unit simplex, so $w_{i,t}\geq 0$ and $\sum_{i=0}^{N}w_{i,t}=1$. The semantic certificate later acts only on the risky subvector $w_t^r=(w_{1,t},\ldots,w_{N,t})^\top$.

Cash receives its own action coordinate because fully invested equity policies have no internal way to lower market exposure when text signals contraction. In the empirical environment, cash is modeled as a liquid zero-return reserve. Increasing $w_{0,t}$ therefore preserves the wealth accounting identity while reducing exposure to the risky factor vector. The gate can use this coordinate without creating short positions or external hedges.

Let $r_{t+1}\in\mathbb{R}^{N+1}$ denote next-period asset returns, with the cash component set to zero unless stated otherwise. The drifted pre-trade portfolio inherited from the previous execution is $\tilde{w}_{t-1}$. With transaction cost rate $c$, the realized net return is
\begin{equation}
R_{p,t+1}=w_t^\top r_{t+1}-c\|w_t-\tilde{w}_{t-1}\|_1 .
\end{equation}
The $\ell_1$ turnover term makes reallocation costly and discourages unnecessary churning. Wealth follows $V_{t+1}=V_t(1+R_{p,t+1})$, so the effect of a safety intervention appears through both exposure and trading friction.

\subsection{LLM-derived semantic risk state}
\label{subsec_semantic_state}

Daily document blocks $D_t$ contain only information available before the allocation for date $t$ is executed. The open-source language model reads this block under a fixed schema and returns a JSON state $z_t$ rather than a trade direction. The continuous fields include severity $\mathrm{sev}_t$, uncertainty $\mathrm{unc}_t$, macroeconomic stress $\mathrm{mac}_t$, liquidity stress $\mathrm{liq}_t$, earnings stress $\mathrm{earn}_t$, and model confidence $\mathrm{conf}_t$. Categorical fields record regimes such as $\texttt{risk\_off}$ and $\texttt{risk\_on}$, together with affected-sector scores $a_{i,t}\in[0,1]$.

The numerical state used by the certificate is $x_t=(\mathrm{sev}_t,\mathrm{unc}_t,\mathrm{mac}_t,\mathrm{liq}_t,\mathrm{earn}_t,1-\mathrm{conf}_t)^\top$. Broad textual stress is
\begin{equation}
s_t=\mathrm{clip}(\langle \omega^m,x_t\rangle,0,1),
\end{equation}
where $\omega^m$ is fixed across all folds. A declared $\texttt{risk\_off}$ regime raises $s_t$ to at least $0.60$, whereas a $\texttt{risk\_on}$ regime attenuates the broad stress score. This scalar controls the market-wide part of the adverse center.

Sector exposure enters through the affected-sector intensity
\begin{equation}
g_{i,t}=(\beta_0+\beta_e\mathrm{earn}_t+\beta_s\mathrm{sev}_t+\beta_u\mathrm{unc}_t)a_{i,t} .
\end{equation}
The semantic shock center for risky asset $i$ is
\begin{equation}
m_{i,t}^{\mathrm{sem}}=-\lambda_m\delta_i(s_t+g_{i,t}) .
\end{equation}
The negative sign converts adverse text into a conservative return-shock center rather than a return forecast. The attenuation $\delta_i$ reduces broad-stress loading for Consumer Staples, Utilities, and Healthcare, while nondefensive sectors retain unit loading.

Textual stress also widens the dispersion assigned to exposed sectors. Define $b_t=\langle \omega^\phi,x_t\rangle$ and
\begin{equation}
\psi_{i,t}=\mathrm{clip}\!\left(\delta_i'\left[b_t+(\beta_0'+\beta_e'\mathrm{earn}_t)a_{i,t}+\mathbb{1}_{i\in\{\mathrm{XLE},\mathrm{XLF}\}}\beta_{\mathrm{EF}}(\mathrm{mac}_t+\mathrm{liq}_t)\right],0,2\right).
\end{equation}
Energy and Financials receive the extra macro-liquidity loading because commodity-price, credit, and funding channels make these sectors more exposed to systemic stress. The diagonal inflation coefficient is
\begin{equation}
\phi_{i,t}^{\mathrm{sem}}=1+\lambda_\phi\psi_{i,t} .
\end{equation}

The semantic-to-geometry mapping is treated as a frozen domain-prior configuration rather than as a fold-level search target. Its monotone structure follows financial text evidence and sector-risk reasoning \cite{LoughranMcDonald2011}. Higher severity, uncertainty, macro stress, liquidity strain, and low confidence can only increase adverse center shifts or dispersion. Defensive sectors receive weaker broad-stress loading, while Energy and Financials receive additional macro-liquidity exposure. All constants are fixed before the reported walk-forward evaluation and remain unchanged across training, validation, and test folds. They are not re-estimated on test windows or optimized by regime.

\subsection{Semantic uncertainty set}
\label{subsec_semantic_uncertainty_set}

The rolling covariance estimator $\widehat{\Sigma}_t$ is computed from the 252 trading days preceding date $t$ over risky ETF returns. Semantic dispersion rescales this baseline through $D_t=\mathrm{diag}(\sqrt{\phi_{1,t}^{\mathrm{sem}}},\ldots,\sqrt{\phi_{N,t}^{\mathrm{sem}}})$, producing
\begin{equation}
\Sigma_t^{\mathrm{sem}}=D_t\widehat{\Sigma}_tD_t+\epsilon I,
\qquad \epsilon=10^{-6} .
\end{equation}
The ridge term keeps the shape matrix invertible. The uncertainty set is the ellipsoid
\begin{equation}
\mathcal{U}_t(\rho_t)=\left\{u\in\mathbb{R}^N\,\middle|\,(u-m_t^{\mathrm{sem}})^\top(\Sigma_t^{\mathrm{sem}})^{-1}(u-m_t^{\mathrm{sem}})\leq \rho_t^2\right\} .
\end{equation}
A rectangular set would treat sector shocks as separable marginal deviations. The ellipsoid instead preserves cross-sector co-movement through $\Sigma_t^{\mathrm{sem}}$, while the semantic inflation terms resize directions exposed by the current text.

\subsection{Conformal calibration of semantic risk}
\label{subsec_conformal_calibration}

Model confidence does not by itself provide calibrated coverage for future factor shocks, since neural classifiers often assign excessive probability to wrong outputs \cite{Guo2017}. Gated-CSRC therefore measures the realized distance between the semantic certificate and the next risky return vector. For each historical date with observable $r_{t+1}^r$, the nonconformity score is
\begin{equation}
e_t=\sqrt{(r_{t+1}^r-m_t^{\mathrm{sem}})^\top(\Sigma_t^{\mathrm{sem}})^{-1}(r_{t+1}^r-m_t^{\mathrm{sem}})} .
\end{equation}
The score records how far the realized return lies from the text-conditioned center under the semantic shape.

Global exchangeability is not plausible across long financial samples that mix calm trading, liquidity contractions, policy shocks, and crash periods. The calibration design therefore compares the current certificate with historical dates assigned to the same semantic bucket, following the conformal principle that empirical residuals should determine prediction-set size \cite{vovk2005algorithmic,ShaferVovk2008}. This local comparison does not remove non-stationarity. It restricts the calibration sample to dates with similar text-derived stress states.

The bucket rule assigns each date to one category through a fixed priority order. Low-confidence or high-uncertainty days enter the first bucket when $\mathrm{conf}_t<0.35$ or $\mathrm{unc}_t\geq0.75$. Liquidity stress, macroeconomic stress, and earnings stress then receive priority when their corresponding field exceeds $0.65$. Broad risk-off days follow when severity exceeds $0.70$ or the regime field declares $\texttt{risk\_off}$. Low-severity $\texttt{risk\_on}$ dates enter a calm bucket. Remaining dates are neutral.

Let $b_t$ denote the active semantic bucket and let $\mathcal{C}_t=\{t-504,\ldots,t-1\}$ denote the rolling two-year calibration window. If the bucket contains $n$ matching scores, the candidate radius is the empirical order statistic
\begin{equation}
\rho_t=Q_{1-\alpha}\left(\{e_\tau\,|\,\tau\in\mathcal{C}_t,\ b_\tau=b_t\}\right),
\end{equation}
with rank $\lceil(n+1)(1-\alpha)\rceil$ after sorting the matching scores in ascending order. The radius is always computed from past dates only.

Sparse buckets create noisy quantiles. When fewer than 60 matching observations exist, the method falls back to the unconditional rolling history over $\mathcal{C}_t$, provided at least 126 historical scores are available. This fallback favors operational stability over a thin regime-specific estimate, since rare text states can otherwise produce unstable radii.

The target miscoverage level is fixed at $\alpha=0.10$, so the procedure targets 90\% empirical coverage within the rolling bucketed calibration design. A physical floor $\rho_{\min}=0.005$ prevents zero-volume sets during calm document periods. The executed radius is $\rho_t\leftarrow\max(\rho_t,\rho_{\min})$.

\subsection{Worst-case semantic loss evaluation}
\label{subsec_worst_case_loss}

The calibrated ellipsoid becomes actionable only after it is evaluated against a portfolio. For the risky subvector $w_t^r$, define the semantic loss as the worst adverse portfolio return over the uncertainty set,
\begin{equation}
\mathcal{L}_t(w_t^r)=\sup_{u\in\mathcal{U}_t(\rho_t)}\left(-(w_t^r)^\top u\right) .
\end{equation}
This scalar is the pre-trade quantity checked by the gate.

The supremum has a closed-form dual representation because $\mathcal{U}_t(\rho_t)$ is ellipsoidal. For any candidate risky allocation,
\begin{equation}
\mathcal{L}_t(w_t^r)=-(w_t^r)^\top m_t^{\mathrm{sem}}+\rho_t\sqrt{(w_t^r)^\top\Sigma_t^{\mathrm{sem}}w_t^r} .
\end{equation}
The first term measures exposure to the semantic shock center. The second term scales the portfolio standard deviation under the text-conditioned shape matrix by the conformal radius.

This expression avoids a nested adversarial optimization problem inside the execution loop. Each candidate allocation requires a quadratic-form evaluation under $\Sigma_t^{\mathrm{sem}}$, which is computationally small for the fixed nine-ETF universe used in the experiments. The gate can therefore score a proposal before invoking any projection solver.

Cash sits outside the risky semantic factor vector. It has no semantic center shift and no semantic dispersion loading in $\mathcal{L}_t(w_t^r)$. Raising $w_{0,t}$ reduces risky weights through the simplex identity and lowers exposure to the uncertainty set. The loss function therefore gives the projection layer a direct numerical reason to move capital into liquid reserves during severe document states.

\subsection{Gated semantic safety projection layer}
\label{subsec:gated_projection}

The PPO actor emits a raw proposal $\tilde{w}_t$ on the long-only simplex. A deterministic preprocessing step applies the semantic cash floor and the training-time turnover limiter, giving the pre-check proposal $\bar{w}_t$. The gate evaluates $\mathcal{L}_t(\bar{w}_t^r)$ against the active budget $l_t$. If the proposal is admissible, execution uses $w_t=\bar{w}_t$ and no conic program is called.

The budget is tied to reference portfolios rather than to a static numerical ceiling. Let
\begin{equation}
q_t=\mathrm{clip}(\langle\omega^l,x_t\rangle,0,1)
\end{equation}
measure textual stress for loss-budget tightening. Define
\begin{equation}
L_t^{\mathrm{ref}}=\min\left(\mathcal{L}_t((w^{\mathrm{eq}})^r),1.08\mathcal{L}_t((w^{\mathrm{def}})^r)\right),
\end{equation}
where $w^{\mathrm{eq}}$ is the equal-weight risky allocation with the configured cash rule and $w^{\mathrm{def}}$ is the defensive allocation specified in the experimental design.

Stress contracts the reference loss through
\begin{equation}
\kappa_t=\mathrm{clip}(0.95-0.25q_t,0.70,0.98),
\end{equation}
and the executable budget is
\begin{equation}
l_t=\max\left(l_{\min},\kappa_tL_t^{\mathrm{ref}},1.03\mathcal{L}_t(w_{t-1}^r)\right),
\qquad l_{\min}=0.020 .
\end{equation}
The inherited-portfolio term keeps yesterday's executed allocation near the feasible set after sudden textual shocks. The gate fires only when $\mathcal{L}_t(\bar{w}_t^r)>l_t$.

A fired gate solves a second-order cone projection for the executable allocation. The objective keeps the corrected portfolio close to the policy proposal while adding a weak anchor to inherited holdings,
\begin{equation}
w_t=\arg\min_{w}\frac{1}{2}\|w-\bar{w}_t\|_2^2+\frac{\eta}{2}\|w-w_{t-1}\|_2^2,
\qquad \eta=10^{-3} .
\end{equation}
The anchor limits abrupt liquidation when a single news day produces a temporary certificate expansion.

The feasible set enforces long-only exposure, turnover control, volatility control, and the semantic loss bound,
\begin{equation}
\begin{aligned}
&\sum_{i=0}^{N}w_{i,t}=1,\qquad w_{i,t}\geq0,\\
&\|w_t-w_{t-1}\|_1\leq\tau_{\max},\\
&\sqrt{(w_t^r)^\top\widehat{\Sigma}_t w_t^r}\leq \sigma_{\max}/\sqrt{252},\qquad \sigma_{\max}=0.25,\\
&-(w_t^r)^\top m_t^{\mathrm{sem}}+\rho_t\sqrt{(w_t^r)^\top\Sigma_t^{\mathrm{sem}}w_t^r}\leq l_t .
\end{aligned}
\end{equation}
Sector exposure enters this system through $m_t^{\mathrm{sem}}$ and $\Sigma_t^{\mathrm{sem}}$ rather than through separate linear sector caps.

The projection uses Clarabel as the primary conic solver, with ECOS and SCS fallback attempts if numerical failure occurs. If the exact problem is infeasible, the relaxation ladder widens the semantic budget by factors of $1.25$, $1.50$, and $2.00$ while loosening turnover and volatility limits in the same sequence. Persistent infeasibility triggers a deterministic defensive allocation over cash, Consumer Staples, Utilities, and Healthcare.

The semantic cash floor operates before the gate. Define
\begin{equation}
\tilde{q}_t=\mathrm{clip}(\langle\omega^c,x_t\rangle,0,1),
\qquad c_t^{\mathrm{floor}}=c_{\max}\tilde{q}_t^{p} .
\end{equation}
Macro, liquidity, and broad risk-off buckets lift the floor to at least $0.20$, while the low-confidence bucket lifts it to at least $0.25$. When $\tilde{w}_{0,t}$ is below the floor, risky weights are scaled down proportionally until cash reaches $c_t^{\mathrm{floor}}$. The resulting $\bar{w}_t$ is the only proposal evaluated by the gate. Algorithm~\ref{alg_gated_csrc_execution} summarizes the full pre-trade execution sequence.

\begin{algorithm}[t]
\caption{Gated-CSRC pre-trade execution}
\label{alg_gated_csrc_execution}
\begin{algorithmic}[1]
\State Observe price features, semantic state $x_t$, and prior allocation $w_{t-1}$
\State PPO proposes raw allocation $\tilde{w}_t$
\State Apply cash floor and turnover preprocessing to obtain $\bar{w}_t$
\State Construct $m_t^{\mathrm{sem}}$, $\Sigma_t^{\mathrm{sem}}$, $\rho_t$, and $l_t$
\If{$\mathcal{L}_t(\bar{w}_t^r)\leq l_t$}
    \State Execute $w_t=\bar{w}_t$
\Else
    \State Solve the SOCP projection for $w_t$
    \If{the projection is infeasible}
        \State Apply the relaxation ladder and then the defensive fallback if needed
    \EndIf
\EndIf
\State Record source documents, semantic state, certificate, intervention flag, weights, and constraint residuals
\end{algorithmic}
\end{algorithm}

\subsection{Policy optimization and reward formulation}
\label{subsec_rl_policy}

The actor observes market history, semantic scalars, and inherited holdings. At date $t$, $o_t$ stacks a standardized lookback window of price-derived features for the risky ETFs, the six continuous semantic fields, and the previously executed allocation $w_{t-1}$. Fold-level scalers are fit only on the training window and then reused out of sample. An actor-critic network maps $o_t$ to allocation logits, and a softmax gives the long-only proposal $\tilde{w}_t=\pi_\theta(o_t)$ over risky ETFs and cash. Proximal policy optimization updates $\theta$ on rolling training windows \cite{Schulman2017}. The actor sees $x_t$ but not $m_t^{\mathrm{sem}}$, $\Sigma_t^{\mathrm{sem}}$, or $\rho_t$ directly, and training does not solve the projection problem inside the gradient loop.

The step reward combines cost-aware log growth with penalties for turnover and semantic-budget breaches,
\begin{equation}
r_{t+1}=\kappa_R\ln(1+R_{p,t+1})-\lambda_{\mathrm{turn}}\|\bar{w}_t-w_{t-1}\|_1-\lambda_{\mathrm{sem}}\max\left(0,\mathcal{L}_t(\bar{w}_t^r)-l_t\right) .
\end{equation}
The reported experiments use $\kappa_R=100$, $\lambda_{\mathrm{turn}}=0.02$, and $\lambda_{\mathrm{sem}}=20$. Penalizing the pre-projection proposal teaches the actor to avoid gate activation during training, while the execution gate enforces the boundary when the learned proposal still breaches the certificate.

\section{Experimental setup}
\label{sec:experimental-setup}

The empirical platform synchronizes daily sector-ETF execution, timestamped financial text, transaction costs, semantic state extraction, conformal calibration, and projection-based portfolio control within one chronological backtesting environment.

\subsection{Data sources and asset universe}
\label{subsec:data-universe}

The risky universe contains nine U.S. Select Sector SPDR ETFs with complete coverage over the sample period. The instruments are Materials (XLB), Energy (XLE), Financials (XLF), Industrials (XLI), Technology (XLK), Consumer Staples (XLP), Utilities (XLU), Consumer Discretionary (XLY), and Healthcare (XLV). Real Estate and Communication Services are excluded because they do not satisfy the same full-horizon availability requirement. The resulting cross-section preserves broad macroeconomic sector exposure without mixing assets whose histories begin under different data regimes.

Daily adjusted closing prices span January 1, 2015, through December 31, 2023. The same calendar anchors the textual stream drawn from the Financial News and Stock Price Integration Dataset, which provides time-aligned financial news and stock records over a broad 1999--2023 horizon \cite{dong2024fnspid}. The experiment uses the shared 2015--2023 window so that price returns and text states enter the decision process at the same daily frequency.

Daily frequency still requires a precise publication boundary. Each article is assigned to a trading date according to its timestamp. Any article released after the official market close is shifted to the next eligible trading date. The semantic state available for allocation at date $t$ therefore contains only text that would have been observable before the corresponding portfolio decision. This rule applies before feature construction, LLM inference, conformal scoring, and policy execution.

\subsection{Financial text processing and LLM inference}
\label{subsec:llm-inference}

For each trading date, the preprocessing routine forms one textual block by concatenating exact article headlines with the first 200 characters of each article body. When the number of available documents exceeds the context budget, the daily sequence is capped at 20 documents.

The cap keeps inference feasible while retaining a cross-section of the day's financial narrative. Structured extraction then uses the 8-billion-parameter Llama-3-Instruct model, loaded through the Hugging Face \texttt{transformers} interface \cite{grattafiori2024llama3,wolf2020transformers}. The model is executed with 4-bit weight quantization and half-precision computation to reduce memory demand during offline batch inference \cite{dettmers2023qlora}. Quantization is treated as an engineering constraint, not as evidence of unchanged extraction accuracy.

The decoding temperature is fixed at zero. This disables stochastic sampling and turns the language model into a deterministic schema extractor for the same input block and execution environment. The parsed state vector contains continuous variables in $[0,1]$ for severity, uncertainty, liquidity risk, macro risk, earnings risk, confidence, and affected-sector intensities. The actor receives these scalars as observations, whereas the certificate construction in Section~\ref{sec:methodology} maps them into loss geometry. The language model does not receive authority to rank assets or propose trades.

Schema failures are handled before the simulation advances. Missing fields, invalid numeric ranges, markdown wrappers, and malformed JSON trigger a logged neutral fallback state rather than a terminated backtest. The fallback does not assert that market risk is absent. It prevents a formatting error from becoming an execution failure in the sequential experiment.

\subsection{Prompt engineering and schema enforcement}
\label{subsec:prompt-engineering}

Conversational generation conflicts with automated portfolio control because explanatory prose breaks downstream parsers. The system prompt assigns a quantitative extraction role, prohibits non-JSON output, and defines \texttt{risk\_on}, \texttt{neutral}, and \texttt{risk\_off} regimes. Routine positive news maps to near-zero severity. Negative market states require broad panic, severe macro shocks, or multi-sector contagion.

The user prompt injects the trading date and the assembled text block, then closes with a fixed JSON schema specifying field names and numeric ranges. The schema forces typed semantic variables for liquidity, macro policy, earnings, confidence, and sector channels. Table~\ref{tab:prompt-design} records the prompt template used during extraction.

\begin{table}[h]
\centering
\caption{System and user prompt architecture for daily semantic extraction.}
\label{tab:prompt-design}
\footnotesize
\begin{tabular}{p{0.95\linewidth}}
\toprule
\textbf{System Prompt} \\
\midrule
You are a quantitative financial sentiment analyzer. Your task is to assess the overall market regime and specific sector risks based ONLY on the provided news.\\
Rules:\\
1. \texttt{risk\_on} means bullish and optimistic. Use this or \texttt{neutral} if news is positive or routine, and set severity close to 0.0.\\
2. \texttt{risk\_off} means bearish, panic, or crisis. Only use \texttt{risk\_off} if there is widespread panic, severe macro shocks, or multi-sector contagion.\\
3. Return ONLY valid JSON. No markdown formatting, no explanations. \\
\midrule
\textbf{Dynamic User Prompt Template} \\
\midrule
Date: \{date\_str\} \\
Market News: \{news\_text\} \\
Return JSON with fields: \\
\{ \\
\texttt{"regime"}: \texttt{"risk\_on | neutral | risk\_off"}, \\
\texttt{"event\_types"}: [\texttt{"macro"}, \texttt{"earnings"}, \texttt{"liquidity"}, \texttt{"geopolitical"}, \texttt{"policy"}, \texttt{"sector"}], \\
\texttt{"severity"}: float in [0.0, 1.0], \\
\texttt{"uncertainty"}: float in [0.0, 1.0], \\
\texttt{"liquidity\_risk"}: float in [0.0, 1.0], \\
\texttt{"macro\_risk"}: float in [0.0, 1.0], \\
\texttt{"earnings\_risk"}: float in [0.0, 1.0], \\
\texttt{"affected\_sectors"}: mapping from sector labels to floats in [0.0, 1.0], \\
\texttt{"confidence"}: float in [0.0, 1.0] \\
\} \\
\bottomrule
\end{tabular}
\end{table}

\subsection{Walk-forward validation protocol}
\label{subsec:walk-forward}

Evaluation follows rolling walk-forward partitions. Each fold uses a 48-month training window followed by a 12-month test window. The fold sequence is 2015--2018 training for 2019, 2016--2019 training for 2020, 2017--2020 training for 2021, 2018--2021 training for 2022, and 2019--2022 training for 2023. The out-of-sample period therefore includes the 2020 crash, the policy response that followed it, and the subsequent recovery under changing inflation and rate expectations.

The training window fixes the learned and assigned objects that could otherwise absorb test information. PPO weights, feature scalers, reward constants, semantic transformation constants, and hyperparameter choices are determined within the 48-month block and then kept unchanged during the corresponding test year. Test-window returns, losses, and benchmark outcomes do not revise these quantities.

Rolling statistical objects follow the information set available at the decision date. When covariance estimates and conformal radii update during testing, they use only observations dated before the allocation under evaluation. This convention preserves the sequential timing of the method while separating causal updating from test-period parameter selection. A fold can therefore adapt to realized history inside the test year only after that history has become observable.

\subsection{Trading parameters and hyperparameters}
\label{subsec:trading-parameters}

The market simulator charges proportional transaction costs to make rebalancing costly. The cost rate $c$ is fixed at 10 basis points per unit of traded portfolio weight. Turnover is the two-way $\ell_1$ change $\|w_t-w_{t-1}\|_1$. Raw PPO proposals are capped at $0.35$ per risky asset before the gate, while the projection solver can move up to $0.50$ when feasibility requires stronger repositioning. These limits separate the actor's ordinary allocation range from the solver's corrective authority.

PPO training uses a learning rate of $1\times 10^{-4}$, rollout length of 256, batch size of 64, five epochs per update, discount factor $\gamma_{\mathrm{RL}}=0.99$, generalized advantage parameter $0.95$, and $50{,}000$ environment interactions per fold \cite{Schulman2017}. Table~\ref{tab:hyperparameters} reports the full policy, reward, certificate, and projection settings.

\begin{table}[h]
\centering
\caption{Policy, reward, certificate, and projection hyperparameters.}
\label{tab:hyperparameters}
\begin{tabular}{lll}
\toprule
Group & Parameter & Value \\
\midrule
PPO & Learning rate & $1\times10^{-4}$ \\
PPO & Rollout length $n_{\text{steps}}$ & 256 \\
PPO & Batch size & 64 \\
PPO & Epochs per update & 5 \\
PPO & Discount $\gamma_{\text{RL}}$ & 0.99 \\
PPO & GAE $\lambda$ & 0.95 \\
PPO & Clip range & 0.10 \\
PPO & Entropy coefficient & 0.003 \\
PPO & Value coefficient & 0.50 \\
PPO & Max gradient norm & 0.50 \\
PPO & Target KL & 0.02 \\
PPO & Hidden layers actor / critic & $[128, 64]$, Tanh \\
PPO & Interactions per fold & 50{,}000 \\
Reward & Return scale $\kappa_R$ & 100 \\
Reward & Turnover weight $\lambda_{\text{turn}}$ & 0.02 \\
Reward & Semantic weight $\lambda_{\text{sem}}$ & 20 \\
Certificate & Center scale $\lambda_m$ & 0.020 \\
Certificate & Shape scale $\lambda_\phi$ & 2.0 \\
Certificate & Ridge $\epsilon$ & $10^{-6}$ \\
Certificate & Radius floor $\rho_{\min}$ & 0.005 \\
Projection & Turnover cap proposal / solver & 0.35 / 0.50 \\
Projection & Volatility cap annual & 0.25 \\
Projection & Prior anchor $\eta$ & $10^{-3}$ \\
Projection & Budget floor $l_{\min}$ & 0.020 \\
\bottomrule
\end{tabular}
\end{table}

The semantic certificate uses a conformal miscoverage level of $\alpha=0.10$, giving a nominal 90\% coverage target under the local calibration convention. Nonconformity scores are drawn from a two-year rolling window of 504 trading days. Bucket-specific calibration is used when at least 60 observations are available, and the global fallback requires 126 observations. The radius floor is $\rho_{\min}=0.005$.

The geometry constants in Table~\ref{tab:certificate_config} define how semantic fields move the center, shape, loss budget, and cash floor. The center scale is $\lambda_m=0.020$, the shape scale is $\lambda_\phi=2.0$, and the ridge term is $\epsilon=10^{-6}$. The dynamic semantic budget contracts toward the reference loss in Section~\ref{subsec:gated_projection} and is bounded below by $l_{\min}=0.020$. The annualized volatility cap is $0.25$, and the projection objective retains a prior-portfolio anchor of $\eta=10^{-3}$.

The semantic transformation vectors $(\omega^m,\omega^\phi,\omega^l,\omega^c,\beta,\delta)$ are treated as a fixed monotone domain-prior configuration rather than fold-level estimands. Their signs and relative ordering reflect financial lexicon evidence and sector-risk reasoning \cite{LoughranMcDonald2011}. Higher severity, uncertainty, macro stress, liquidity strain, and lower confidence may move adverse centers, enlarge uncertainty geometry, tighten the execution budget, or raise the cash floor. Defensive sectors receive weaker broad-stress loading, while Energy and Financials receive additional macro-liquidity exposure.

Retrospective grid search over certificate geometry would weaken the separation between semantic risk control and performance fitting. The transformation constants are therefore held invariant across all walk-forward folds. Sensitivity analysis varies interpretable control dimensions around this canonical configuration, especially conformal coverage $\alpha$ and budget aggressiveness, rather than refitting hidden spatial weights to historical outcomes.

\begin{table}[h]
\centering
\caption{Semantic certificate configuration. Field-weight vectors are listed in the order of $x_t=(\text{sev},\,\text{unc},\,\text{mac},\,\text{liq},\,\text{earn},\,1-\text{conf})$. The constants define a fixed monotone domain-prior configuration and are held invariant across all walk-forward folds.}
\label{tab:certificate_config}
\begin{tabular}{lll}
\toprule
Symbol & Role & Value \\
\midrule
$\omega^{m}$ & Center severity weights & $(0.35,\,0.15,\,0.25,\,0.20,\,0,\,0.05)$ \\
$\omega^{\phi}$ & Shape inflation weights & $(0.30,\,0.25,\,0.20,\,0.15,\,0,\,0.10)$ \\
$\omega^{l}$ & Budget stress weights & $(0.35,\,0.25,\,0.20,\,0.20,\,0,\,0)$ \\
$\omega^{c}$ & Cash-floor stress weights & $(0.30,\,0.20,\,0.20,\,0.15,\,0.10,\,0.05)$ \\
$(\beta_0,\beta_e,\beta_s,\beta_u)$ & Center sector loadings & $(0.55,\,0.25,\,0.10,\,0.10)$ \\
$(\beta_0',\beta_e')$ & Shape sector loadings & $(0.70,\,0.30)$ \\
$\beta_{\text{EF}}$ & Energy/financials loading & $0.20$ \\
$\delta_i$ & Defensive attenuation center & $0.75$ \\
$\delta_i'$ & Defensive attenuation shape & $0.85$ \\
\bottomrule
\end{tabular}
\end{table}

\subsection{Benchmark models}
\label{subsec:baselines}

Comparative evaluation uses classical allocation rules and learning-based alternatives under identical price histories, text inputs where applicable, transaction costs, and walk-forward folds. Equal-weight allocation is included because naive diversification is difficult to beat out of sample \cite{demiguel2009optimal}. Momentum rules are included because sector rotation often begins from recent relative strength signals \cite{jegadeesh1993returns}. Each benchmark is evaluated after the same calendar alignment and cost accounting used for Gated-CSRC.

The Equal Weight benchmark holds $1/N$ of risky capital in each of the nine ETFs and no cash. The Inverse Volatility benchmark scales each ETF by the reciprocal of its trailing 20-day return standard deviation and remains fully invested. Momentum Top-3 ranks sectors by 60-day momentum divided by 20-day volatility, then allocates only across the three highest-scoring ETFs.

The Momentum LegacyCap variant adds an uncalibrated semantic ceiling before projection. The ceiling for asset $i$ is $0.35(1-0.75\,\mathrm{risk}_{i,t})$, floored at $0.05$, with excess capital redistributed to assets that retain spare capacity. Raw PPO receives the same semantic risk scalars as state variables but executes without the CSRC projection gate, isolating text-conditioned policy learning from certificate-based execution control.

The Hard CSRC baseline applies the projection layer every trading day regardless of semantic trigger intensity. This comparator measures the cost of continuous mathematical intervention, including turnover and transaction-cost drag. The benchmark set therefore separates static diversification, volatility scaling, price momentum, heuristic semantic caps, text-conditioned reinforcement learning, unconditional projection, and gated projection.

\section{Empirical Results}
\label{sec:results}

The execution architecture is evaluated as a risk-constraint mechanism rather than as a stand-alone forecasting device. The experiments separate financial return preservation from semantic limit compliance and certificate-conditioned execution stability. Language-model-derived news states therefore enter the evaluation as calibrated geometric inputs that can block unsafe allocations before execution, rather than as uncalibrated return predictors.

\subsection{Classical benchmark evaluation}
\label{subsec_benchmark_evaluation}

The equal-weight portfolio establishes a demanding static control because naive diversification often remains difficult to beat after estimation error, turnover, and sampling noise are considered \cite{demiguel2009optimal}. In the present out-of-sample window, the $1/N$ allocation achieves a $12.06\%$ annual return and a $0.670$ Sharpe ratio. It also records no semantic violations without any projection. That zero-breach result matters because it prevents an inflated interpretation of CSRC. The layer should be judged as a safety governor that corrects unsafe proposals, not as a mechanism that must dominate every static allocation rule.

That safety-governor interpretation becomes sharper under volatility scaling. The inverse-volatility rule lowers concentration mechanically and reaches a $10.87\%$ annual return with a $0.631$ Sharpe ratio, yet it still breaches the semantic certificate on $0.20\%$ of trading days. Adding the continuous CSRC projection removes these minor breaches, while the Sortino ratio moves from $0.735$ to $0.710$ as feasibility enforcement absorbs a small execution cost.

Momentum exposes the opposite failure mode. The Top-3 rule follows winner persistence in the spirit of classical relative-strength allocation \cite{jegadeesh1993returns}, but the concentrated rebalancing path generates a $23.48\%$ semantic violation rate and a $-44.95\%$ maximum drawdown. Projection suppresses those breaches to $0.10\%$, although the portfolio still pays substantial turnover and constraint-adjustment costs. The result is informative because the same safety layer that is nearly redundant for equal weight becomes essential once the baseline concentrates exposure in recent winners. Table~\ref{tab_main_benchmark_comparison} reports the full static benchmark comparison.

\begin{table}[h]
\centering
\caption{Main out-of-sample benchmark comparison over the 2019--2023 test period.}
\label{tab_main_benchmark_comparison}
\footnotesize
\setlength{\tabcolsep}{2.7pt}
\begin{tabular}{@{}lrrrrrrr@{}}
\toprule
 & Ann. & Ann. &  &  & Max & Avg. & Semantic \\
Strategy & return & vol. & Sharpe & Sortino & drawdown & turnover & violation \\
\midrule
Equal Weight & 12.06\% & 18.00\% & 0.670 & 0.791 & -36.93\% & 0.56\% & 0.00\% \\
Equal Weight + CSRC & 11.15\% & 17.41\% & 0.640 & 0.749 & -37.36\% & 0.91\% & 0.00\% \\
Inverse Volatility & 10.87\% & 17.22\% & 0.631 & 0.735 & -35.98\% & 3.28\% & 0.20\% \\
Inverse Volatility + CSRC & 10.36\% & 16.88\% & 0.614 & 0.710 & -36.43\% & 3.48\% & 0.00\% \\
Momentum Top-3 & -0.12\% & 20.20\% & -0.006 & -0.007 & -44.95\% & 33.81\% & 23.48\% \\
Momentum Top-3 + CSRC & 0.94\% & 17.98\% & 0.052 & 0.062 & -41.70\% & 25.53\% & 0.10\% \\
Momentum Top-3 LegacyCap + CSRC & 4.84\% & 16.91\% & 0.286 & 0.336 & -35.65\% & 18.74\% & 0.00\% \\
\bottomrule
\end{tabular}
\end{table}

\subsection{Reinforcement learning policy and semantic safety trade-offs}
\label{subsec_ppo_csrc}

The static benchmark results leave the main question at the execution boundary, where a learned policy can discover profitable allocation paths while still proposing certificate-infeasible trades. Raw PPO therefore serves as the return-seeking proposal engine in the central comparison \cite{Schulman2017}. The term Raw PPO refers to execution without the projection layer, rather than to a training process stripped of semantic inputs. It reaches a $12.7\%$ mean annual return and a $1.13$ Sharpe ratio, indicating that the neural actor extracts economically useful allocation structure from the training environment.

That financial strength does not carry its own semantic protection. Raw PPO violates the time-aligned language-model-derived risk certificate on 51 of 1,256 out-of-sample trading days, approximately $4.06\%$ of the paired evaluation sample. Averaged across the five annual test folds, the corresponding rate is $4.05\%$, which is the figure reported in Table~\ref{tab_ppo_csrc_comparison} alongside the fold-averaged return and Sharpe ratio. The extra return relative to safer variants therefore arrives with an uncompensated semantic tail-exposure channel. Table~\ref{tab_ppo_csrc_comparison} gives the policy-level comparison.

The unconditional Hard CSRC variant resolves that exposure through feasibility at every decision time. It records an observed $0.00\%$ semantic violation rate under strict walk-forward certificate scoring, yet the constant projection lowers mean annual return to $10.9\%$ and the Sharpe ratio to $1.04$. The loss is not a numerical accident. An always-on projector can alter allocations even when the current certificate already permits the raw proposal, so safe execution is purchased partly through unnecessary correction during benign certificate states.

Gated CSRC changes the intervention rule rather than the feasibility definition. The gate leaves Raw PPO untouched when the proposed allocation satisfies the current semantic budget and solves the projection only after a certificate breach is detected. This state-contingent execution path also records an observed $0.00\%$ semantic violation rate, while recovering mean annual return to $11.7\%$ and the Sharpe ratio to $1.09$. The comparison with Hard CSRC isolates the value of gating itself, since both variants enforce the same certificate but differ in how often they invoke the solver.

\begin{figure}[h]
    \centering
    \includegraphics[width=1\textwidth, height=0.4\textheight, keepaspectratio]{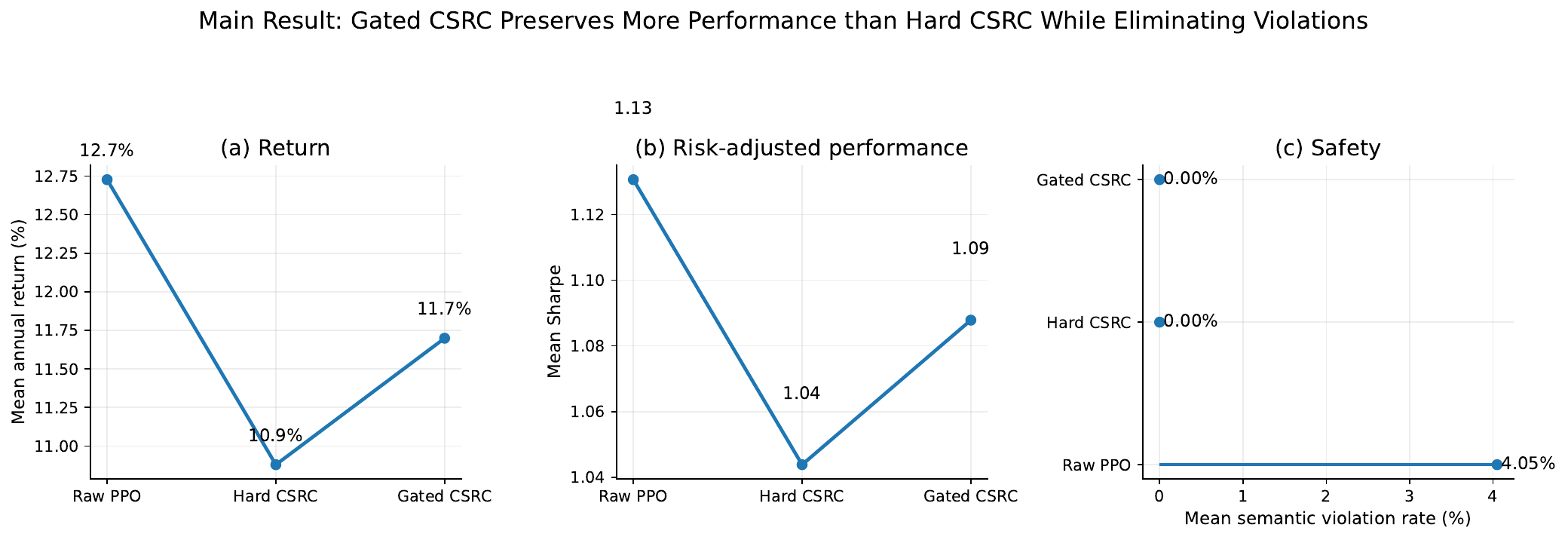}
    \caption{Main CSRC result. Raw PPO delivers competitive performance but violates the semantic risk certificate. Hard CSRC eliminates observed violations with a larger performance cost. Gated CSRC eliminates observed violations while recovering more of the risk-adjusted return path.}
    \label{fig_main_csrc_result}
\end{figure}

Figure~\ref{fig_main_csrc_result} makes the same trade-off visible in the executed return paths. Raw PPO remains financially effective but semantically exposed, whereas Hard CSRC removes the exposure with a larger performance cost. Gated CSRC preserves more of the native return-seeking path because the correction is concentrated on certificate-breach days and leaves feasible proposals unchanged.

\begin{table}[h]
\centering
\caption{Raw PPO versus hard and gated semantic risk control. All entries are averages over the five annual test folds.}
\label{tab_ppo_csrc_comparison}
\begin{tabular}{lccc}
\toprule
Strategy & Mean annual return & Mean Sharpe & Mean semantic violation rate \\
\midrule
Raw PPO & 12.7\% & 1.13 & 4.05\% \\
Hard CSRC & 10.9\% & 1.04 & 0.00\% \\
Gated CSRC & 11.7\% & 1.09 & 0.00\% \\
\bottomrule
\end{tabular}
\end{table}

\subsection{Selective intervention across chronological regimes}
\label{subsec_gated_intervention_results}

The state-contingent rule should intervene most often when the news certificate describes acute stress. Chronological fold diagnostics support that pattern. During the 2020 systemic liquidity crisis, elevated severity vectors and tighter semantic budgets cause Gated CSRC to intercept $77.9\%$ of Raw PPO proposals, while still recording a $0.00\%$ final violation rate. This behavior distinguishes the gate from a calendar-based de-risking rule. The intervention rate rises because the certificate geometry tightens under contemporaneous textual stress, not because the year is treated as an ex ante crisis label. Table~\ref{tab_gated_intervention_by_year} reports the year-level diagnostics.

\begin{figure}[h]
    \centering
    \includegraphics[width=1.0\textwidth, height=0.4\textheight, keepaspectratio]{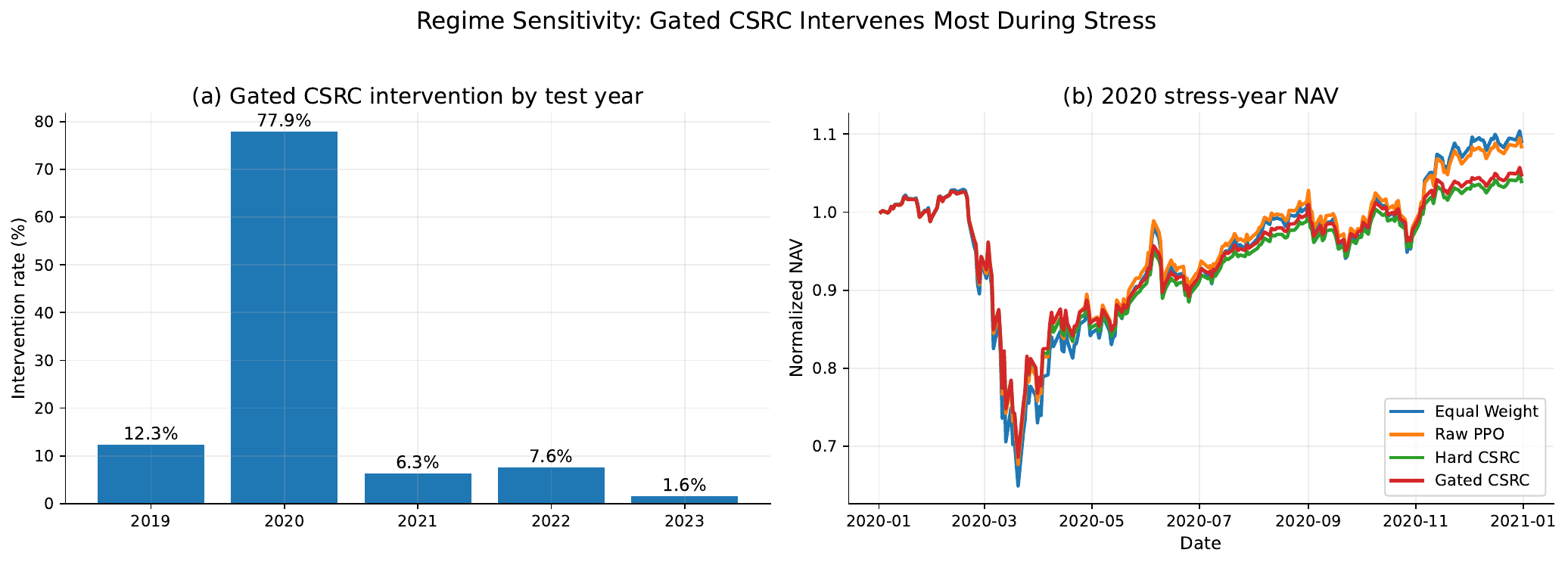}
    \caption{Selective intervention and stress-year behavior. Panel (a) reports the Gated-CSRC intervention rate by test year. Panel (b) compares normalized NAV paths during the 2020 stress year.}
    \label{fig_intervention_and_nav}
\end{figure}

The same mechanism becomes nearly dormant when certificate stress subsides. In the 2023 testing fold, intervention frequency falls to $1.6\%$, which indicates that the solver does not operate as a permanent conservative overlay. Figure~\ref{fig_intervention_and_nav} places intervention rates beside normalized net asset value paths, showing that most mathematical correction occurs when the semantic boundary is actively threatened.

\begin{table}[h]
\centering
\caption{Gated-CSRC intervention diagnostics by test year.}
\label{tab_gated_intervention_by_year}
\begin{tabular}{lccc}
\toprule
Test year & Intervention rate & Final violation rate & Interpretation \\
\midrule
2019 & 12.3\% & 0.00\% & Moderate intervention \\
2020 & 77.9\% & 0.00\% & High intervention during stress \\
2021 & 6.3\% & 0.00\% & Low intervention \\
2022 & 7.6\% & 0.00\% & Low-to-moderate intervention \\
2023 & 1.6\% & 0.00\% & Minimal intervention \\
\bottomrule
\end{tabular}
\end{table}

\subsection{Semantic ablation of the language model certificate}
\label{subsec_semantic_ablation_results}

The intervention results depend on the certificate being both textual and time aligned. The ablation study replaces the language-model-derived shock vectors with static, shuffled, and dictionary-based controls while preserving the same projection geometry. Each substitute certificate passes through the same execution pipeline, and final violations are scored against the true time-aligned LLM certificate. This design keeps the test focused on semantic information rather than on solver differences, because the optimization layer receives the same mathematical form in every ablation.

Neutral and shuffled certificates fail because they remove distinct pieces of the risk state. The Neutral control collapses every semantic field to zero across all dates, creating a static reference state with no event response. The Time-Shuffled control preserves the marginal distribution of LLM states but breaks alignment between news and market dates, producing a $4.05\%$ final violation rate.

The Loughran--McDonald certificate fails for a different reason \cite{LoughranMcDonald2011}. It counts category token rates and converts them into causal expanding percentiles, yet dictionary counts cannot represent issuer roles, event timing, sector interactions, or contagion context. A sentence about supply-chain disruption and a sentence about ordinary uncertainty can therefore share token categories while implying different portfolio hazards. Its executed allocations therefore leak $4.45\%$ final true violations under the same strict scoring rule, as shown in Table~\ref{tab_semantic_ablation_summary}.

\begin{figure}[h]
    \centering
    \includegraphics[width=0.95\textwidth, height=0.4\textheight, keepaspectratio]{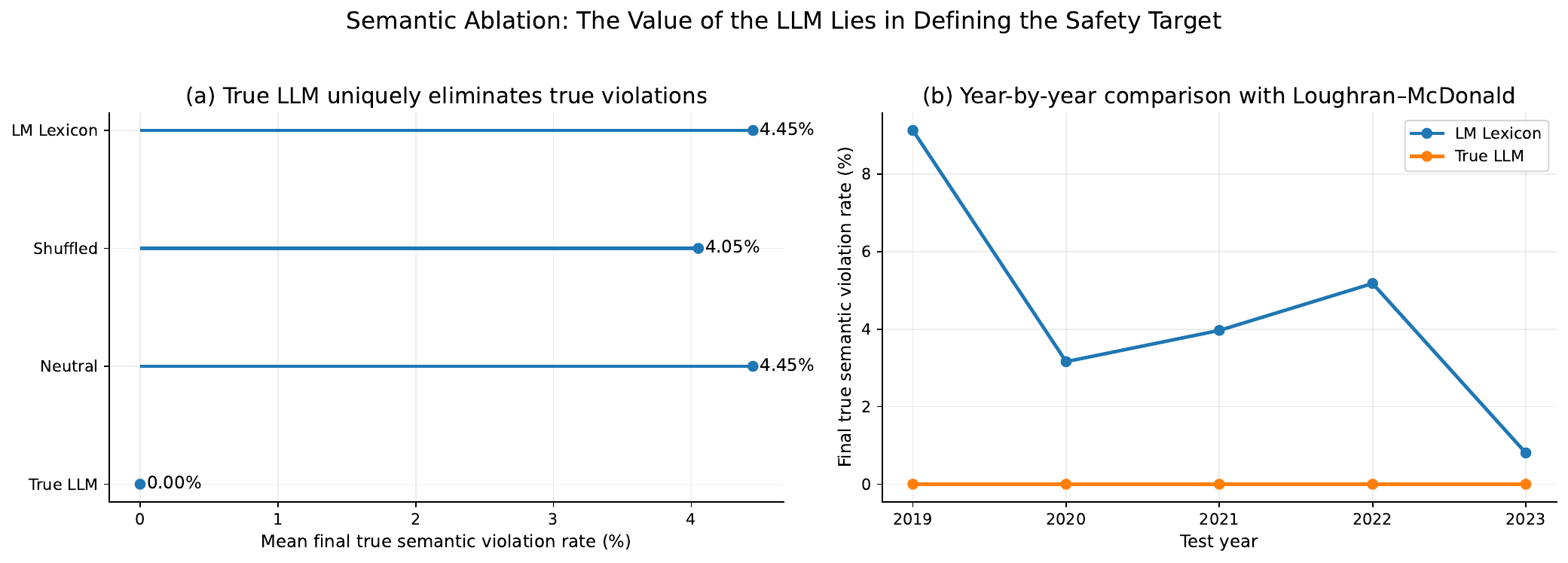}
    \caption{Semantic ablation. Panel (a) reports mean final true semantic violation rates across certificate variants. Panel (b) compares the true LLM certificate against the Loughran--McDonald lexicon baseline by test year.}
    \label{fig_semantic_ablation}
\end{figure}

Figure~\ref{fig_semantic_ablation} links these failures to the certificate source. The True LLM variant records an observed $0.00\%$ violation rate, while Neutral, Shuffled, and lexicon controls remain exposed under the true certificate. The evidence supports context-sensitive semantic extraction as a risk-certificate generator rather than a direct trading signal, with temporal alignment carrying measurable execution value.

\begin{table}[h]
\centering
\caption{Semantic ablation results. Final violations are measured under the true time-aligned LLM semantic certificate.}
\label{tab_semantic_ablation_summary}
\begin{tabular}{lccc}
\toprule
Semantic certificate & Mean annual return & Mean Sharpe & Final true violation rate \\
\midrule
True LLM & 11.7\% & 1.09 & 0.00\% \\
Neutral & 11.8\% & 1.09 & 4.45\% \\
Shuffled LLM & 11.6\% & 1.08 & 4.05\% \\
Loughran--McDonald lexicon & 11.7\% & 1.09 & 4.45\% \\
\bottomrule
\end{tabular}
\end{table}

\subsection{Geometric risk-aversion sensitivity}
\label{subsec_risk_aversion_results}

Risk-aversion profiles expose how much return the execution layer can preserve after the institution fixes its admissible semantic exposure. Table~\ref{tab_risk_profiles} changes the conformal miscoverage level, cash ceiling, cash exponent, macro floor, and gate tolerance while holding the budget floor fixed. The resulting executions are evaluated under a strict benchmark based on the conservative certificate. The comparison therefore treats profile relaxation as a conscious policy choice, not as a change in the definition of safety after the fact.

\begin{table}[h]
\centering
\caption{Risk-aversion profile parameters.}
\label{tab_risk_profiles}
\begin{tabular}{lccccc}
\toprule
Profile & $\alpha$ (coverage) & Cash ceiling & Cash exponent & Macro floor & Gate tolerance \\
\midrule
Conservative & 0.10 (90\%) & 0.25 & 2.25 & 0.20 & $10^{-8}$ \\
Moderate & 0.15 (85\%) & 0.20 & 2.10 & 0.15 & $2.5\times10^{-3}$ \\
Aggressive & 0.25 (75\%) & 0.15 & 2.00 & 0.10 & $5\times10^{-3}$ \\
\bottomrule
\end{tabular}
\end{table}

Strict true violation scoring prevents relaxed profiles from being evaluated against their own weaker budgets. For every profile, a violation is counted when the executed portfolio exceeds the conservative semantic loss budget. Under that standard, the Conservative profile intercepts $21.1\%$ of policy proposals and records an observed $0.00\%$ true semantic violation rate.

The Moderate profile expands the feasible region by loosening empirical coverage and gate parameters. Intervention frequency falls to $15.0\%$, so fewer Raw PPO proposals receive mathematical correction. The execution path gains financial room, raising mean annual return to $11.8\%$ and the Sharpe ratio to $1.10$ under the reported profile setting.

That added room also admits unsafe actions under the strict benchmark. When Moderate executions are rescored against the conservative certificate, true semantic violations reappear on $3.7\%$ of evaluated trading days. The profile therefore does not fail its own relaxed design; it demonstrates the cost of moving the safety boundary outward.

The Aggressive profile gives the return-seeking actor still more freedom. In the evaluated folds, it produces no solver activations, so the reported intervention rate is $0.0\%$. This path reaches the highest financial metrics in the profile study, with a $12.9\%$ mean annual return and a $1.14$ Sharpe ratio.

Relaxed execution remains exposed even when the violation percentages are not perfectly ordered. Aggressive records a $3.3\%$ strict true violation rate, slightly below Moderate but still far above the Conservative zero. Table~\ref{tab_risk_aversion_summary} therefore supports a configurable return-safety exchange, not a monotone violation law, with realized breaches depending on the specific path of proposed allocations.

\begin{figure}[h]
    \centering
    \includegraphics[width=0.95\textwidth, height=0.4\textheight, keepaspectratio]{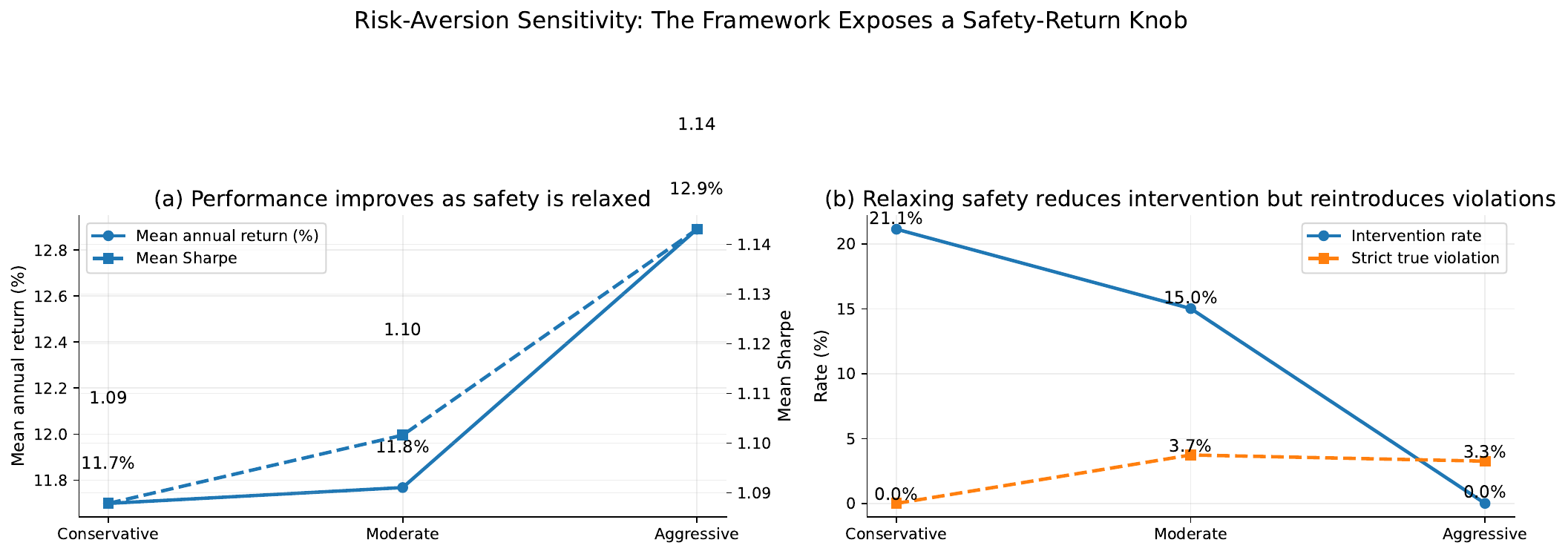}
    \caption{Risk-aversion sensitivity. Panel (a) shows financial performance across profile settings. Panel (b) shows intervention rates and strict true semantic violations under conservative rescoring.}
    \label{fig_risk_aversion}
\end{figure}

Figure~\ref{fig_risk_aversion} places that exchange on the same visual scale. Lower intervention coincides with higher realized performance, but nonzero strict certificate breaches return once the conservative geometry is relaxed. The architecture forces risk tolerance to become an explicit execution parameter before capital is committed, rather than a hidden consequence of reward tuning.

\begin{table}[h]
\centering
\caption{Risk-aversion sensitivity under strict conservative rescoring.}
\label{tab_risk_aversion_summary}
\begin{tabular}{lcccc}
\toprule
 & Mean annual & Mean & Intervention & Strict true \\
Profile & return & Sharpe & rate & violation rate \\
\midrule
Conservative & 11.7\% & 1.09 & 21.1\% & 0.0\% \\
Moderate & 11.8\% & 1.10 & 15.0\% & 3.7\% \\
Aggressive & 12.9\% & 1.14 & 0.0\% & 3.3\% \\
\bottomrule
\end{tabular}
\end{table}

\subsection{Statistical validation of the compliance guarantee}
\label{subsec_statistical_validation}

The violation series provide paired binary evidence because every architecture is evaluated on the same 1,256 out-of-sample trading dates. Raw PPO breaches the true semantic certificate on 51 dates, while Gated CSRC records no such breaches. This paired construction avoids comparing unrelated samples and isolates the execution-layer correction.

Exact McNemar testing applies directly to those paired violation indicators \cite{mcnemar1947note}. The null hypothesis of equal violation frequency is rejected for Gated CSRC versus Raw PPO with $p<10^{-15}$, as Table~\ref{tab_safety_stat_tests} reports. The same rejection holds for Hard CSRC versus Raw PPO, while Gated CSRC and Hard CSRC are indistinguishable on the binary safety outcome.

\begin{table}[h]
\centering
\caption{Exact McNemar hypothesis testing on paired daily violation indicators across execution architectures.}
\label{tab_safety_stat_tests}
\begin{tabular}{lccccc}
\toprule
 & Violation & Violation & A violates, & A safe, &  \\
Comparison & A & B & B safe & B violates & $p$-value \\
\midrule
Gated CSRC vs Raw PPO & 0.00\% & 4.06\% & 0 & 51 & $<10^{-15}$ \\
Hard CSRC vs Raw PPO & 0.00\% & 4.06\% & 0 & 51 & $<10^{-15}$ \\
Gated CSRC vs Hard CSRC & 0.00\% & 0.00\% & 0 & 0 & 1.000 \\
\bottomrule
\end{tabular}
\end{table}

Financial performance tests carry weaker distributional support than the binary safety test. The Jobson--Korkie paired Sharpe comparison with Memmel's correction returns a $p$-value of $0.493$ \cite{jobson1981performance,memmel2003performance}. The statistic is useful because it asks whether the observed Sharpe difference is large relative to paired return variation. Daily portfolio returns exhibit non-normality and serial dependence, however, so the result should be read as an approximate diagnostic rather than as a full distributional statement.

Mean-return differences are checked with a Newey--West test to account for residual autocorrelation and heteroskedasticity \cite{newey1987simple}. With the automatic seven-lag bandwidth used for this sample length, the test returns a $p$-value of $0.175$. The result does not establish return equality; it only fails to detect a statistically measurable mean-return loss.

That distinction keeps the empirical claim inside the design target. Gated CSRC sharply reduces paired semantic violations, while the reported tests do not detect financial underperformance relative to Raw PPO. The evidence therefore supports safety improvement without a return-dominance interpretation, leaving economic superiority outside the tested claim under the sample and metrics used here.

The resulting audit trail has direct value for model-risk review. Each daily record connects a raw allocation, a certificate, a gate decision, and a final violation indicator. Risk managers can reproduce the safety test without interpreting the hidden state of the policy network or the internal activations of the language model. The unit of verification is the replayable execution record.

\subsection{Summary of findings}
\label{subsec_results_summary}

The empirical results support five claims about the proposed architecture. Equal Weight confirms that CSRC is a safety governor rather than a universal return engine. Raw PPO remains financially effective but semantically exposed when execution lacks a certificate projection. Gated CSRC records zero observed strict certificate breaches while preserving more return than an always-on projector. Semantic ablations show that neutral, shuffled, and dictionary-based controls cannot reproduce the time-aligned contextual certificate. Risk-aversion profiles convert tolerance choices into visible return-safety exchange points. The statistical tests reinforce the same separation between safety and return claims, since paired violation reductions are decisive while financial outperformance is not established. Taken together, the evidence supports Gated-CSRC as a mechanism for recovering most of an RL policy's return while enforcing a certified risk bound, with LLM outputs restricted to calibrated risk-certificate construction rather than direct trading authority.

\section{Conclusion}
\label{sec:conclusion}

Gated-CSRC develops an auditable execution layer that links timestamped financial text to continuous reinforcement-learning portfolio control. The language model does not act as a speculative trading oracle. It produces schema-constrained semantic risk states that can be transformed into checkable certificates over factor shocks and portfolio losses. PPO remains responsible for proposing return-seeking weights, whereas conformal calibration and the asymmetric projection gate decide whether those weights can be executed. This separation gives the system the form of a pre-trade review process in which narrative information changes the feasible action set before capital is committed.

The walk-forward evidence over nine U.S. Select Sector SPDR ETFs shows a consistent return-safety pattern. Raw PPO retains strong allocation capacity, yet it exposes the portfolio to strict semantic certificate violations during periods when narrative stress and market dislocation are not absorbed by the raw action proposal. Gated-CSRC reduces the observed strict semantic certificate violation rate to zero in the reported out-of-sample tests while activating the solver selectively rather than replacing the policy at every step. Bag-of-words risk dictionaries and time-shuffled text controls do not reproduce the same bounding behavior, which indicates that timestamped semantic geometry supplies information beyond static lexical tone.

The same evidence fixes the scope of the claim. Direct LLM trading agents solve a signal-generation problem in which text is mapped to discrete actions or point forecasts. Gated-CSRC instead evaluates an execution-feasibility problem after a numerical portfolio policy has already proposed a continuous allocation. This distinction makes LLM trading agents relevant background, but not the primary counterfactual for testing the gate. The present design still depends on local approximate exchangeability of return nonconformity scores inside rolling two-year semantic buckets. It also uses cached daily semantic certificates, which suits institutional end-of-day rebalancing more naturally than ultra-low-latency trading on streaming news.

The certificate-and-gate structure generalizes beyond this application: any setting where a learned policy proposes actions under a calibrated, replayable uncertainty set can adopt the same separation between proposal and certified feasibility. For capital markets practitioners and model-risk managers specifically, the provenance protocol has direct institutional value. Each daily execution record connects a timestamped source document, an extracted semantic state, a calibrated certificate radius, a raw policy proposal, a gate decision, and a final intervention flag. A compliance officer can reconstruct whether a specific allocation satisfied its pre-trade risk constraint without interpreting the internal activations of the language model or the policy network. The 100-basis-point auditable safety premium observed in the walk-forward tests is therefore best read as a measured execution cost of institution-grade AI control in continuous portfolio allocation.

These latency and calibration limits define the next stage of the architecture. Future work can replace single-pass semantic extraction with committee-based certificate validation, so that disagreement among language models becomes an explicit input to the risk set rather than an informal diagnostic. Rolling conformal calibration can also be adapted to higher-frequency news tapes where semantic regimes change faster than the current bucket schedule assumes. Neural alpha generation and mathematical downside control can coexist when executable actions are convexly bounded by replayable certificates, intervention logs, and model-risk evidence.

\backmatter

\section*{Declarations}
\subsection*{Competing Interests}

The authors have no competing interests to declare that are relevant to the content of this article.

\subsection*{Author contribution}

Mohammad Zarei: Conceptualization, Methodology, Software, Data curation, Formal analysis, Validation, Visualization, original draft. Amirhossein MohammadpourAzari: Methodology, Software, Validation, review and editing. Moslem Habibi: Supervision, Conceptualization, Methodology, Project administration.

\subsection*{Funding}

This research did not receive any specific grant from funding agencies in the public, commercial, or not-for-profit sectors.

\subsection*{Data availability}

The replication codebase, including implementation scripts, configuration files, and instructions for reproducing the reported experiments, will be made publicly available after publication at \url{https://github.com/MoZarei-AI/Gated_CSRC}. The study uses publicly available market data and the Financial News and Stock Price Integration Dataset. Processed semantic certificates and derived experimental outputs will be released with the repository to the extent permitted by the licenses and terms governing the underlying data sources.

\bibliography{sn-bibliography}%

\end{document}